\documentclass[a4paper,11pt]{article}
\pdfoutput=1
\usepackage{slashed}
\usepackage[affil-sl,auth-sc]{authblk}
\usepackage{amssymb,amsmath,amsbsy}
\usepackage{mathrsfs}
\usepackage{mathbbol}
\usepackage{bm}
\usepackage{appendix}
\usepackage{hyperref}
\usepackage[top=1.2in, bottom=1.2in,left=0.9in,includefoot]{geometry}               
\usepackage{graphicx}
\usepackage{cite}
\usepackage{float}
\usepackage{caption}
\usepackage{wasysym}
\usepackage{float}
\usepackage{caption}
\usepackage[utf8]{inputenc}

\usepackage{subcaption}

\usepackage{amsthm}

\newif\ifContLineOne
\newif\ifContLineTwo
\newif\ifContLineThree

\def\conC#1{\vbox{\ialign{##\crcr
  \ifContLineThree\hrulefill\else\vphantom{\hrulefill}\fi\crcr
  \noalign{\kern3.2pt\nointerlineskip}
  \ifContLineTwo\hrulefill\else\vphantom{\hrulefill}\fi\crcr
  \noalign{\kern3.2pt\nointerlineskip}
  \ifContLineOne\hrulefill\else\vphantom{\hrulefill}\fi\crcr
  \noalign{\nointerlineskip}
  $\hfil\textstyle{\vbox to 14pt{}#1}\hfil$\crcr}}}

\def\DrawLeg#1#2{
  \kern-.2pt              
  \dimen2 =#1             
  \advance\dimen2 by 2pt  
  \dimen3 = 10.6pt        
  \dimen4 =3.6pt          
  \advance\dimen3 by -\dimen2 
  \multiply\dimen4 by #2
  \advance\dimen3 by \dimen4
  \raise\dimen2 \hbox{\vrule height\dimen3 width .4pt} 
  \kern-.2pt}             

\def\begC#1#2{\setbox0 =\hbox{$\textstyle{#2}$}
  \dimen0=.5\wd0 \dimen1=\ht0
  \conC{\hskip\dimen0}
  \count255=#1
  \ifnum\count255 =1 \ContLineOnetrue\else
  \ifnum\count255 =2 \ContLineTwotrue\else
  \ifnum\count255 =3 \ContLineThreetrue\fi\fi\fi
  \DrawLeg{\dimen1}{\count255}
  \conC{\hskip\dimen0}
  \kern-\dimen0\kern-\dimen0 \box0}

\def\endC#1#2{\setbox0 =\hbox{$\textstyle{#2}$}
  \dimen0=.5\wd0 \dimen1=\ht0
  \conC{\hskip\dimen0}
  \count255=#1
  \ifnum\count255 =1 \ContLineOnefalse\else
  \ifnum\count255 =2 \ContLineTwofalse\else
  \ifnum\count255 =3 \ContLineThreefalse\fi\fi\fi
  \DrawLeg{\dimen1}{\count255}
  \conC{\hskip\dimen0}
  \kern-\dimen0\kern-\dimen0 \box0}

\def\eq#1{eq.~(\ref{#1})}

\newcommand{\bea}{\begin{eqnarray}}
\newcommand{\eea}{\end{eqnarray}}

\newcommand{\GeV}{\; \mathrm{GeV}}
\newcommand{\TeV}{\; \mathrm{TeV}}

\usepackage{color}

\definecolor{orange}{rgb}{0.9,0.2,0}
\definecolor{brown}{rgb}{0.7,0.3,0.2}
\definecolor{fuxia}{rgb}{1,0,1}
\definecolor{skyblue}{rgb}{0,0.1,0.9}
\definecolor{violetred}{rgb}{0.8,0.13,0.56}
\definecolor{deeppink}{rgb}{1.00,0.08,0.5}
\definecolor{pink}{rgb}{1.00,0.75,0.80}
\definecolor{orchid}{rgb}{0.85,0.44,0.84}
\definecolor{lightpink}{rgb}{1.00,0.71,0.76}
\definecolor{bluish}{rgb}{0,0.6,0.8}

\title{\vspace{-2cm}
{\small \begin{flushright}
MAN/HEP/2017/06\\[1mm]
April 2017
\end{flushright} }
{\bf Radiative Light Dark Matter}}

\author{

A.~Dedes$^{1}$\footnote{email: {\tt adedes@cc.uoi.gr}}, ~D.~Karamitros$^{1}$\footnote{email: {\tt dkaramit@cc.uoi.gr}}
 ~and A.~Pilaftsis$^{2}$\footnote{email: {\tt Apostolos.Pilaftsis@manchester.ac.uk}}  }
\affil{\small $^{1}$Department of Physics, Division of Theoretical
  Physics, \\ University of Ioannina, GR 45110, Greece}
\affil{\small $^{2}$ Consortium for Fundamental Physics, School of Physics and Astronomy, \\ 
University of Manchester, Manchester, M13 9PL, United Kingdom}
\begin{document}

\maketitle

\begin{abstract}
  We present a Peccei--Quinn (PQ)-symmetric two-Higgs doublet model
  that naturally predicts a fermionic singlet dark matter in the mass
  range 10~keV--1~GeV.  The origin of the smallness of the mass of
  this light singlet fermion arises predominantly at the one-loop
  level, upon soft or spontaneous breakdown of the PQ symmetry via a
  complex scalar field in a fashion similar to the so-called
  Dine--Fischler--Sredniki--Zhitnitsky axion model. The mass
  generation of this fermionic Radiative Light Dark Matter (RLDM)
  requires the existence of two heavy vector-like SU(2) isodoublets,
  which are not charged under the PQ symmetry. We show how the RLDM
  can be produced via the freeze-in mechanism, thus accounting for the
  missing matter in the Universe. Finally, we briefly discuss possible
  theoretical and phenomenological implications of the RLDM model for
  the strong CP problem and the CERN Large Hadron~Collider~(LHC).
\end{abstract}

\bigskip\bigskip

\section{Introduction}
\setcounter{equation}{0}\label{sec:intro}

Ongoing searches for the elusive missing matter component of the
Universe, the so-called Dark Matter (DM), have offered no conclusive
evidence so far. From analyses of the CMB power spectrum and from pertinent
astronomical studies, we now know that about one quarter of the energy
budget of our Universe should be in the form of DM, and so many candidate
theories have been put forward to address this well-known DM problem~\cite{Bertone:2004pz}.
Among the suggested scenarios, those
predicting Weakly Interactive Massive Particles (WIMPs) constitute one
class of popular models that may not only account for the DM itself, but also
leave their footprints in low-energy experiments, or even at
high-energy colliders, such as the LHC \cite{Kahlhoefer:2017dnp}. 
In particular, for WIMPs near the
electroweak scale, the WIMP-nucleon scattering cross section is
estimated to be somewhat below $10^{-46}~\mathrm{cm}^2$ as measured by LUX~\cite{Akerib:2016vxi}.

Projected experiments that lie not very far ahead in future will be
capable of reaching sensitivity in the ballpark
$10^{-47}$--$10^{-48}~\mathrm{cm}^2$~\cite{Aprile:2015uzo}, and so
they will be getting closer to the neutrino-nucleon background cross
section, the infamous ``neutrino floor,'' where disentangling neutrino
signals from those of WIMPs will become almost an impossible
task~\cite{Billard:2013qya}. Therefore, DM models have to be
constructed (or revisited) to avoid such severe constraints, e.g.~by
contemplating scenarios that either sufficiently suppress the
WIMP-nucleon interaction,
or move the DM mass to the sub-GeV or ultra-TeV region.

Several models have been proposed featuring a light DM in the mass
range~$\mathcal{O}$(keV)--$\mathcal{O}$(GeV), such as sterile neutrino
DM~\cite{Asaka:2005an,Kusenko:2009up,Mavromatos:2012cc,Adhikari:2016bei,Heurtier:2016iac},
light scalar DM~\cite{Boehm:2003hm} and milli-charged
DM~\cite{Huh:2007zw}, including their possible implications for future
DM searches~\cite{Essig:2013lka,Alexander:2016aln}.  However, one
central problem of such models is the actual origin of the small mass
for the light DM, which could be more than six orders of magnitude
below the electroweak scale.

In this paper we address this mass hierarchy problem, by presenting a
new radiative mechanism that can predominantly account for the
smallness in mass for the light DM. The so-generated Radiative Light
Dark Matter (RLDM) is a fermionic singlet~$S$ and can naturally
acquire a mass in the desired range: 10~keV--1~GeV. A minimal
realization of this radiative mechanism requires the extension of the
Standard Model (SM) by one extra scalar doublet, resulting in a Peccei--Quinn
(PQ)-symmetric two-Higgs doublet model~\cite{Peccei:1977ur,Peccei:1977hh}, 
augmented by two fermionic heavy vector-like SU(2)
isodoublets $D_1$ and $D_2$, which are not charged under the PQ
symmetry. The mass of the RLDM is predominantly generated at the one-loop
level, upon soft or spontaneous breakdown of the PQ symmetry via a
complex scalar field, e.g.~$\Sigma$, in close analogy to the so-called
Dine--Fischler--Sredniki--Zhitnitsky~(DFSZ) axion model that
addresses the strong CP problem~\cite{Zhitnitsky:1980tq,Dine:1981rt}.

We analyse the production mechanisms of the RLDM in the early
Universe, and show that it can account for its missing matter
component via the so-called freeze-in mechanism~\cite{Hall:2009bx}. In
fact, we illustrate how the freeze-in mechanism remains effective in
the RLDM model, without the need to resort to suppressed Yukawa
couplings. In this context, we investigate two possible scenarios of
both theoretical and phenomenological interest. In the first scenario,
we consider the breaking of the PQ scale~$f_{\rm PQ}$ to be comparable
to the one required for the DFSZ model to solve the strong CP problem,
i.e.~$f_{\rm PQ} \sim 10^{9}$~GeV. We find that such PQ scale can
exist within this realization, provided an appropriate isodoublet
mass~$M_D$ and reheating temperature~$T_{\rm RH}$ is considered.  In
the second scenario, we relax the constraint of the strong CP problem
on~$f_{\rm PQ}$, and investigate its possible lower limit, with the
only requirement that $T_{\rm RH}$ be larger than the critical
temperature~$T_C$ of the SM electroweak phase transition, thus
allowing for the $B+L$-violating sphaleron processes to be in thermal
equilibrium. This requirement is introduced here, so as to leave open
the possibility of explaining the cosmological baryon-to-photon ratio
$\eta_B$ via low-scale baryogenesis mechanisms, such as electroweak
baryogenesis~\cite{Kuzmin:1985mm,Cohen:1993nk} and resonant
leptogenesis~\cite{Pilaftsis:1997jf,Pilaftsis:2003gt,Dev:2014laa,Achelashvili:2016trx}.  
In this second scenario, we find that the heavy Higgs bosons of the  two-Higgs doublet model (2HDM)
may have masses as low as a few~TeV, which are well within reach of
the LHC.

The layout of the paper is as follows. In Section~\ref{sec:Mechanism},
we first introduce the PQ-symmetric 2HDM, augmented with a singlet
fermion~$S$ and a fermionic pair of vector-like doublets~$D_{1,2}$.
Then, we describe the radiative mechanism for the RLDM, once the PQ
symmetry is broken softly, and show that a radiative mass in the
range 10~keV--1~GeV can be naturally generated. In
Section~\ref{sec:DMabundance}, we outline the relevant Boltzmann
equation for computing the relic abundance of the RLDM.  Utilising
the freeze-in mechanism, we present in Section~\ref{sec:results}
numerical estimates for the allowed parameter space of our RLDM
model. Based on these results, we explore the possibility whether our
model can account for the strong CP problem within a scenario similar
to the DFSZ axion model.  Moreover, we investigate whether an absolute
lower limit exists for the heavy Higgs-boson masses in our effective
2HDM. Indeed, we find that our RLDM model may allow for heavy Higgs
bosons at the TeV scale, whose existence can be probed at the LHC.
Finally, Section~\ref{sec:Concl} summarises our conclusions and
outlines possible new directions for further research.

\section{Radiative Mechanism}
\label{sec:Mechanism}

In this section we present a minimal extension of the SM, in which the
small mass of the light DM, in the region
$10~\mathrm{keV}$--$1~\mathrm{GeV}$, can have a radiative origin,
generated at the one-loop level. This radiative mechanism is minimally
realised within the context of a constrained 2HDM
obeying a Peccei--Quinn symmetry. In addition, the model under study
contains a singlet fermion $S$ charged under the PQ symmetry and a fermionic
pair of massive isodoublets $D_{1,2}$ with zero PQ charges.  Finally,
we delineate the parameter space for which a viable scenario of
Radiative Light Dark Matter can be obtained consistent with the
observed relic abundance.

\subsection{The Model}

In the 2HDM under consideration, we impose a global PQ symmetry
$U(1)_{\rm PQ}$, which forbids the appearance of a bare mass term for
the singlet fermion~$S$ at the tree level. This PQ symmetry will be
broken softly or spontaneously which in turn triggers a radiative mass
for~$S$ at the one-loop level. The fermion $S$ is stable and receives
naturally a small sub-GeV mass, leading to a RLDM scenario.  On the
other hand, we note that a candidate for a light DM would probably be
relativistic at its freeze-out, resulting in an extremely large relic
abundance (similar to~\cite{Lee:1977ua}) for the allowed range of DM
masses that are larger than about 3~keV,
e.g.~see~\cite{Irsic:2017ixq,Viel:2013apy}.  Therefore, the DM should
be produced out of thermal equilibrium in the early Universe. The
mechanism that we will be utilising here is the so-called freeze-in
mechanism\cite{Hall:2009bx}, which assumes that the DM particles were
absent initially and are produced only later from the plasma.

\begin{table}[t]
\begin{center}
\begin{tabular}{l||r|r|r|r}
 & $SU(2)_L$ & $U(1)_{Y}$ & $U(1)_{\rm PQ}$ & $\mathbb{Z}_2$ \\
\hline
$S$ & 1~~~~ & 0~~~ & $-1$~~~ & odd \\
\hline
$D_{1}$ & 2~~~~ & $-1$~~~ & 0~~~ & odd \\
\hline
$D_{2}$ & 2~~~~ & 1~~~ & 0~~~  & odd \\
\hline
$\Phi_{1}$ & 2~~~~ & 1~~~ & 1~~~ & even\\
\hline
$\Phi_{2}$ & 2~~~~ & 1~~~ & $-1$~~~ & even
\end{tabular}
\end{center}
\caption{\sf Quantum number assignments of particles
  pertinent to the RLDM Model.} 
\label{charges}
\end{table}

The relevant Yukawa and potential terms of our model are given by
\begin{align}
-\mathcal{L}^{\tiny{Y}} \ =\ & 
Y_1 \epsilon^{ab} \Phi_{1a} \, D_{1b} \, S + Y_2 \Phi^{\dagger a}_{2}
                             \, D_{2a} \, S \, + M_D \epsilon^{ab}
                             D_{1a} \, D_{2b} \, +\
                             \mathrm{H.c.}\;,  \label{UV-Yukawa}\\[2mm] 
V(\Phi_{1},\Phi_{2}) \ =\ & \ m_{11}^2 \Phi^{\dagger \, a}_{1}\Phi_{1a}
                          + m_{22}^2 \Phi^{\dagger \, a}_{2}\Phi_{2a}
                          - m_{12}^2  
(\Phi^{\dagger \, a}_{1}\Phi_{2a} + \mathrm{H.c}) +
                          \dfrac{\lambda_1}{2} (\Phi^{\dagger \,
                          a}_{1}\Phi_{1a})^2\nonumber \\ 
& +\: \dfrac{\lambda_2}{2} (\Phi^{\dagger \, a}_{2}\Phi_{2a})^2 + 
 \lambda_{3} \Phi^{\dagger \, a}_{1}\Phi_{1a}\Phi^{\dagger \,
  b}_{2}\Phi_{2b}+ \lambda_{4} |\Phi^{\dagger \, a}_{1}\Phi_{2 \,a}|^2
  \;, 
 \label{V(phi)}
\end{align}
where $a,b =1,2$ are $SU(2)_L$-group indices (with
$\epsilon^{12}=-\epsilon^{21}=+1$), $S$ is a Weyl-fermion SM singlet,
$D_{1,2}$ are two Weyl-fermion $SU(2)_L$-doublets, and $\Phi_{1,2}$
are two scalar $SU(2)_L$-doublets. A complete list of the PQ and
hypercharge quantum numbers of the aforementioned particles is given
in Table~\ref{charges}, including a $\mathbb{Z}_2$-parity which excludes the
mixing of dark-sector particles with those of the SM. For simplicity,
we assume that the new dark-sector interactions are CP invariant 
and so take their respective couplings to be real in the physical mass
basis.

As can be seen from~\eqref{V(phi)}, we have assumed that the PQ
symmetry is broken by the lowest dimensionally possible mass operator
in the scalar potential~$V$, namely by allowing only the dimension-2 mixing
term~$m^2_{12}$ between $\Phi_{1}$ and $\Phi_{2}$. This dimension-2
operator breaks softly the $U(1)_{\rm PQ}$-symmetry in the potential,
but could result from spontaneous breaking of the $U(1)_{\rm PQ}$ by a
scalar $\Sigma$, which acquires a vacuum expectation value (VEV) 
$\langle \Sigma \rangle \equiv  f_{\rm PQ} \sim m_{12}$ (see section~\ref{sec:results}). If the~PQ-breaking
scale~$f_{\rm PQ}$ is high enough, one may neglect, to a good approximation, the
potential quartic couplings $\lambda_{1,2,3,4}$, as they do not affect
much the radiative mass mechanism and the DM production rates which we
will be discussing in the next section.

The mass parameters $m^2_{11}$ and $m^2_{22}$ of the scalar
potential~$V$ in~\eq{V(phi)} may be eliminated in favour of the
VEVs~$v_{1,2}$ of the Higgs doublets~$\Phi_{1,2}$, by virtue of the
minimization conditions on~$V$ (for a review on 2HDMs, see
\cite{Branco:2011iw}). These VEVs are related to the SM Higgs VEV~$v$,
through: $v^2= v_{1}^2 + v_{2}^2$.  In the kinematic region where
$m_{12}^2\gg v^2$, the mass parameters~$m^2_{11}$ and $m^2_{22}$ are
approximately given by
\begin{align}
\label{eq:m11}
 m_{11}^2 &  \ \approx \ m_{12}^2 \, t_{\beta}  \ + \ \mathcal{O}(v^2)\;, \\
\label{eq:m22}
 m_{22}^2 & \ \approx \ m_{12}^2 \, t_{\beta}^{-1} \ + \ \mathcal{O}(v^2)\;,
\end{align}
where $t_{\beta}\equiv \tan \beta =  v_{2}/v_{1}$.

\subsection{One-Loop Radiative Mass}
\label{sec:radmass}

\begin{figure}[t]
\centering        \includegraphics[width=.65\textwidth]{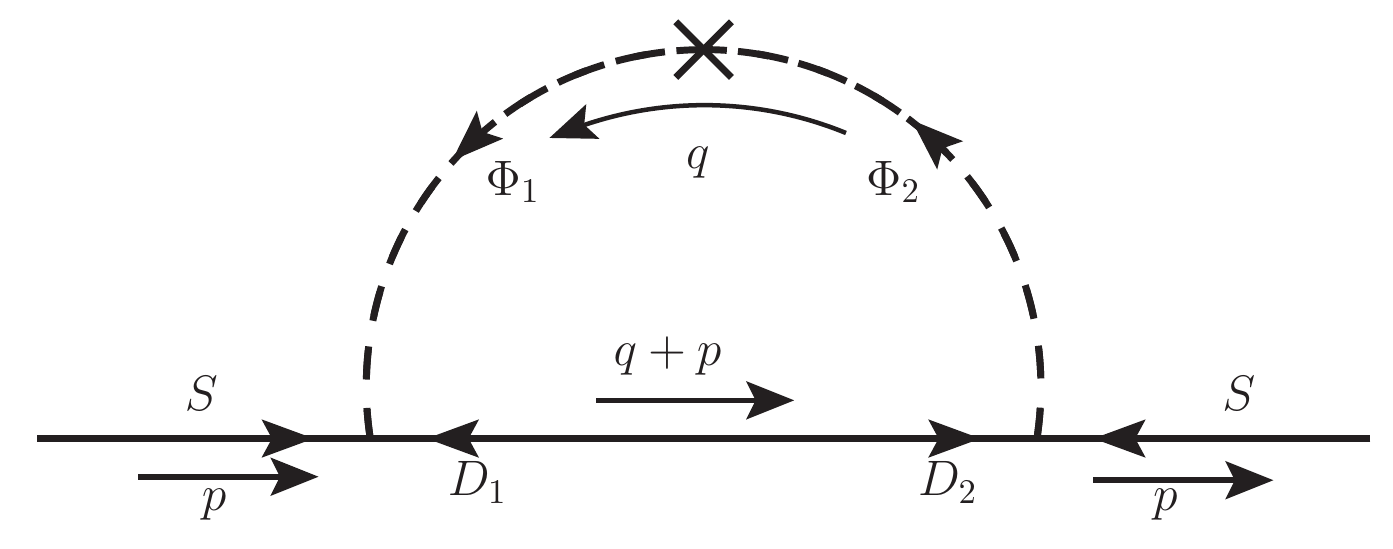}
           \caption{\sf One-loop diagram responsible for the mass
             generation of the singlet fermion $S$.}
           \label{fig:loop}
\end{figure}

Having introduced the minimal model under investigation, we can now
discuss the radiative mechanism responsible for the generation of a
mass of dimension-3 for the singlet fermion~$S$.  We assume that
$m_{12}\stackrel{>}{{}_\sim} 1~\mathrm{TeV}$, such that the main
contribution to the mass of the $S$ particle comes from the diagram
shown in Fig.~\ref{fig:loop}.  In addition, there will be a tree-level mass
$M_S^{\rm tree}$ generated after the SM electroweak phase transition,
given by $M_S^{\rm tree} \simeq Y_{1}Y_{2}\,v^2/M_D$. Under the
assumption that $M_{D}$ is very large, i.e. $M_{D} \gg v$, the tree-level contribution turns out to be
sub-dominant compared to the radiatively induced mass $M_S^{\rm rad}$,
and hence it can be ignored for most of the parameter space. We will
return to this point at the end of this section.

After evaluating the relevant one-loop self-energy graph shown in
Fig.~\ref{fig:loop} at zero external momentum ($p\to 0$), we obtain
\begin{equation}
M_S^{\rm rad}\ =\ -\,2\, Y_{1}Y_{2}\, M_{D}\, m_{12}^2\, I(M_{D},m_{11},m_{22})\;,
\label{MS-loop-diagram}
\end{equation} 
where
\begin{equation}
 I(M_{D},m_{11},m_{22})\ =\ \int\! \dfrac{d^4 q}{(2\pi)^4}\:
 \frac{1}{(q^2 -M_{D}^2) (q^2-m_{11}^2)(q^2-m_{22}^2) }\;. 
\label{MS-loop-I}
\end{equation} 
Employing the approximate relations given in~\eqref{eq:m11} and~\eqref{eq:m22}, the
one-loop radiative mass of~$S$ is finite and may conveniently be expressed as follows:
\begin{equation}
M^{\rm rad}_{S}\ =\ \frac{2 y^2}{(4\pi)^2}\, \frac{M_{D}}{t_{\beta}-t_{\beta}^{-1}} \Bigg[ 
 \frac{t_{\beta} \ln\big(t_{\beta}/r^2\big)}{t_{\beta}-r^2 }\:
-\: \frac{t_{\beta}^{-1} \ln\big(t_{\beta}^{-1}/r^2\big) }{t_{\beta}^{-1}-r^2} \Bigg]\;,
\label{MS-1loop}
\end{equation} 
with $y^2 \equiv Y_{1}Y_{2}$ and $r \equiv M_{D}/m_{12}$.  Observe
that the interchange $t_{\beta} \leftrightarrow t_{\beta}^{-1}$ leaves
$M^{\rm rad}_S$ unchanged.  Assuming that $t_\beta =1$ for different
kinematic regimes of the ratio~$r$, the following simplified forms for
$M^{\rm rad}_S$ are obtained:
\begin{align}
M^{\rm rad}_S\ &\simeq\  \frac{2y^2}{(4\pi)^2} \, M_D \qquad\qquad
                 \mbox{for}\ r\ll 1 \;,  \label{mslimitr=0}\\ 
M^{\rm rad}_S\ &\simeq\  \frac{y^2}{(4\pi)^2} \, M_D  \qquad\qquad
                 \mbox{for}\ r \sim 1 \;,  \label{mslimitr=1}\\ 
M^{\rm rad}_S\ &\simeq\   \frac{2y^2}{(4\pi)^2} \, \frac{M_D\, \ln{r^2}
       }{r^2}\ \ \quad 
\mbox{for}\ r \gg 1 \;. \label{mslimit}
\end{align}
Note that for $M_D\gg m_{12}$ (corresponding to $r\gg 1$), the radiative
mass $M^{\rm rad}_S$ of the singlet fermion~$S$ is suppressed by the
square of the hierarchy factor $r$.  The latter allows for scenarios,
for which the Yukawa couplings are of order~1, i.e.~$y^2 = Y_1 Y_2 =  {\cal O}(1)$,
for $10~\mathrm{keV} \leq M^{\rm rad}_S \leq 1~\mathrm{GeV}$.
On the other hand, for $r\sim 1$ and $r\ll 1$, one needs either a low
$M_D$ of order TeV and $y\approx 0.1$, or
$M_D \approx 10^8$--$10^9~\mathrm{GeV}$ and
$y\approx 10^{-3}$--$10^{-4}$.

\begin{figure}[t]
  \centering
         \includegraphics[width=0.72\textwidth]{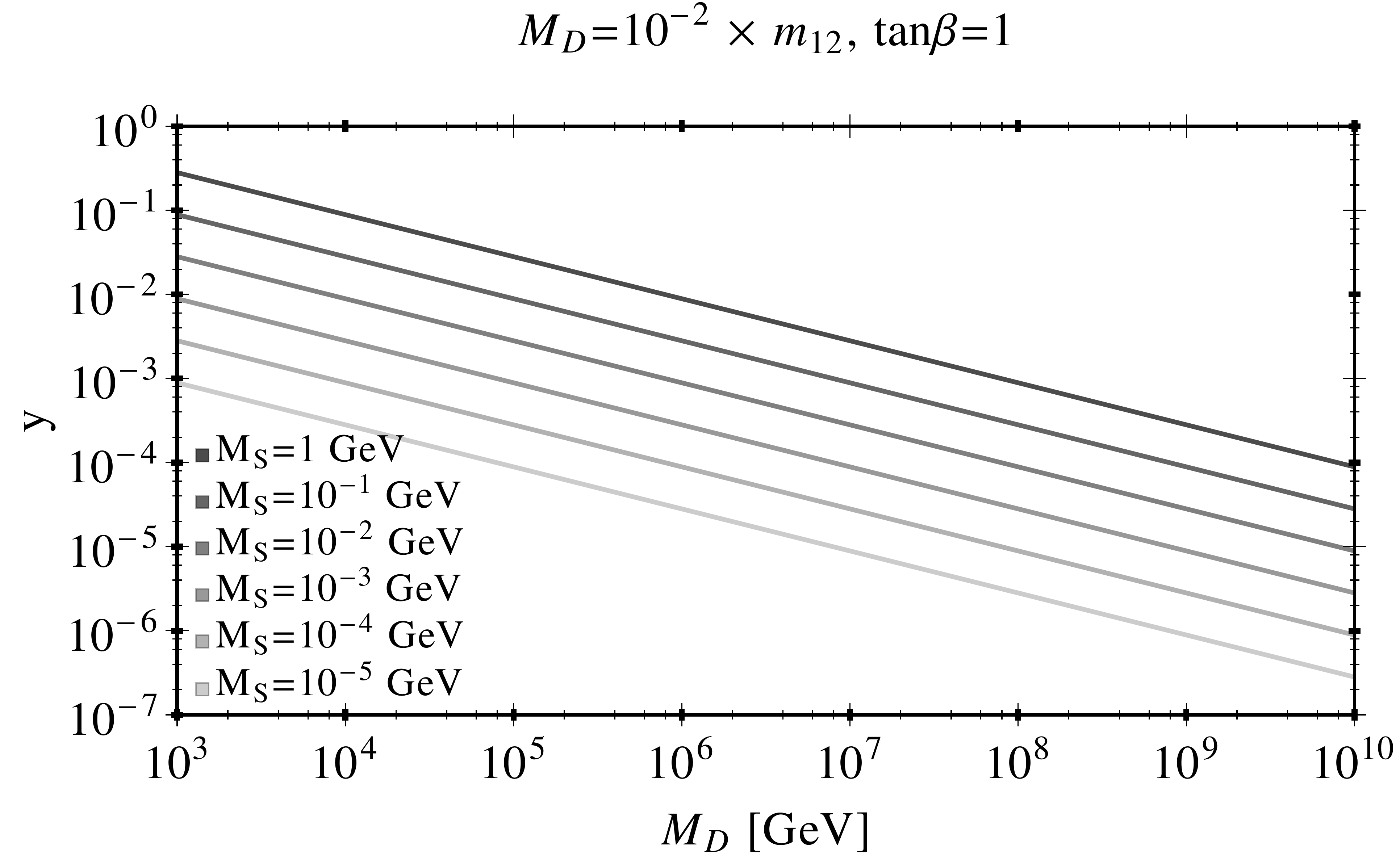}
\caption{\sf Predicted values for $M_D$ versus $y=\sqrt{Y_1 Y_2}$ as
  obtained from 
  \eqref{MS-1loop}, for $M_S \simeq M^{\rm rad}_S$ ranging from $10~\mathrm{keV}$
  to $1~\mathrm{GeV}$,  after setting
     $t_{\beta}=1$ and $r=10^{-2}$. \label{MD-y-MS}}
 \end{figure}

 In Fig.~\ref{MD-y-MS}, we display the values of the coupling
 parameter $y = \sqrt{Y_1Y_2}\,$, as a function of $M_D$, which yield a
 radiatively induced mass $M^{\rm rad}_S$ for the singlet fermion~$S$
 in the region $10~\mathrm{keV} \leq M^{\rm rad}_S \leq 1~\mathrm{GeV}$, for
 $t_{\beta}=1$ and $r=10^{-2}$.  In particular, we see that for every set
 of $M^{\rm rad}_S,\, M_D, \, r$, there is an acceptable range of perturbative
 values for $y$. However, if $r \gg 1$, the desirable value of~$y$ may
 exceed~10 according to~\eqref{mslimit}, and our perturbative results
 do no longer apply. Such non-perturbative values of~$y$ are excluded
from our numerical estimates for the determination of the relic
abundance of~$S$ which we perform in the next section.

In a similar context, we note that a large mass for $m_{11}, m_{12}$, $m_{22}$ and $M_D$ might seem to be a huge fine tuning
for generating a light sub-GeV radiative mass for~$S$. However, we may
easily convince ourselves that this is not the case. The absence of
fine tuning can be seen in an easier way, if we rotate from the
general weak basis spanned by $\Phi_1$ and $\Phi_2$ to the so called
{\it Higgs basis}~\cite{Georgi:1978ri,Branco:2011iw}, $H_1$ and $H_2$, where $H_1$
contains the SM VEV $v$ and $H_2$ has exactly no VEV. Note that in the Higgs
basis, the assignment of the PQ charges to the fields $H_1$ and $H_2$
is not canonical. Moreover, in this rotated Higgs basis, one has that
the new Higgs-mass parameters obey the relation:
$\widetilde{m}_{22}^2\gg \widetilde{m}_{11}^2,\,
\widetilde{m}_{12}^2$. In addition, the
analogue of the diagram in Fig.~\ref{fig:loop} is now represented by a set
of two self-energy graphs, where the fields $H_1$ and $H_2$ are circulating in the
loop. The ultraviolet (UV) infinities cancel, after the
contributions from these two diagrams are added. For $t_\beta=1$ and $r=1$, we
then obtain the same result as the one stated in~\eqref{mslimitr=1}.
Hence, we observe that a small mass for the singlet fermion~$S$ arises
naturally in an SM+$S$ effective field theory. This effective field
theory results from integrating out the heavy $D_{1,2}$ and $H_2$
fields from~\eqref{UV-Yukawa} in the Higgs basis. 

Besides the radiative mass~$M^{\rm rad}_S$ of~$S$ which violates the
PQ symmetry by two units ({\it cf.}~Table~\ref{charges}), there will be a
tree-level contribution to the mass of $S$ after the SM electroweak
phase transition. For most $r$ values of interest here, the relative size of the two
contributions can naively be estimated to be
\begin{equation}
\frac{M_S^{\rm tree}}{M_S^{\rm rad}} \ \sim \  \frac{8 \pi^2 v^2}{M_D^2} \;.
\end{equation}
Thus, for $M_D \gg \sqrt{8} \pi v \simeq 2.2~\mathrm{TeV}$, the
tree-level contribution can be safely ignored. In our numerical 
estimates, the tree-level mass term~$M_S^{\rm tree}$ is always less than 10\% of
the radiative mass term~$M_S^{\rm rad}$. Hence, the total mass~$M_S$ of
the stable fermion $S$ is given predominantly by the radiative mass term,
implying that $M_S\simeq M_S^{\rm rad}$ to a very good approximation. 

We conclude this section by commenting on the possibility of
considering a radiative model alternative to the one discussed here.
For instance, one may envisage a scenario that instead of the single
$S$, one of the neutral components of the doublets $D_{1,2}$ becomes
the RLDM. In~this~case, however, the charged component $D^\pm$ from
$D_{1,2}$ will be almost degenerate with the light sub-GeV DM
particle, which is excluded experimentally.  The general SM+$D_{1,2}$
effective theory has been studied in \cite{Dedes:2016odh}.

\section{Dark Matter Abundance}
\label{sec:DMabundance}

In this section we first describe the relevant effective Lagrangian
that governs the production of the stable fermions $S$ in the early
Universe. We then solve numerically the Boltzmann equation that determines the
yield $Y_S \equiv n_S/s$ of these fermions $S$, where $n_S$ is the
number density of $S$ particles and $s$ is the entropy density of the
plasma.  Having thus estimated the value of $Y_S$, we can then use it to deduce the
respective relic abundance $\Omega_S h^2$ of the $S$ particles in the
present epoch.  Finally, we present approximate analytic results for
$\Omega_S h^2$ and compare these with the observationally favoured
value: $\Omega_{\rm DM} h^2 \simeq 0.12$.

As mentioned in the previous section, the stable fermions $S$ will
play the role of the DM, which are produced via the freeze-in
mechanism\cite{Hall:2009bx}. The key assumption is that the DM
fermions~$S$ were absent (i.e. their number density was suppressed) in
the early Universe and were produced later from annihilations and
decays of plasma particles, e.g.~from $\Phi_{1,2}$ and~$D_{1,2}$,
according to the model discussed in
Section~\ref{sec:Mechanism}. Furthermore, we will assume that
$D_{1,2}$ were also absent in the early Universe, so as to avoid
over-closure of the Universe, unless the Yukawa couplings $Y_{1,2}$
are taken to be extremely suppressed, such that decays of the sort
$D_{1}^{0} \rightarrow h\, S$ are made slow and inefficient. The
latter results in a contrived scenario, in which obtaining a viable DM
parameter space requires a good degree of fine tuning. In order for
the $SU(2)_L$-doublet fermions~$D_{1,2}$ to be absent, we take their
bare mass $M_D$ to be above the reheating
temperature~$T_{\rm RH}$ of the Universe. This simplifies considerably
our analysis, as the heavy fermions $D_{1,2}$ can be integrated out.

The effective Lagrangian that determines the production rate of $S$
particles after reheating is given by
\begin{equation}
    \label{phiphiSS}
-\mathcal{L}_{\rm eff}^{d=5} \ = \ \frac{1}{2\Lambda} \left(
  \Phi_{1}^{\dagger}\Phi_{1}+\tilde{a}
  \Phi_{2}^{\dagger}\Phi_{2}+\tilde{b}\Phi_{1}^{\dagger}\Phi_{2}+\tilde{c}\Phi_{2}^{\dagger}\Phi_{1}
\right) SS\  +\ \mathrm{H.c.}\;, 
\end{equation} 
where $\tilde{a},\tilde{b}$ and $\tilde{c}$ denote the Wilson
coefficients of the dimension-5 operators. The calculation of the
relic abundance is not straightforward in this basis, since
$\Phi_{1,2}$ mix and the identification of the physical fields is
obscured, especially after SSB where further mixing between the scalar
fields is introduced.  Therefore, according to our discussion at the
end of Section~\ref{sec:radmass}, it would be more convenient to
rotate the scalars to the so-called Higgs basis~\cite{Branco:2011iw},
where only one doublet~$H_1$ develops a VEV and is identified with the
SM Higgs doublet. To further simplify calculations, and without much
loss of generality, we assume that the Higgs basis is also the mass
eigenstate basis. This assumption is well justified for relatively
large values of~$m_{12}$, as~it leads to the so-called \emph{alignment
  limit} of the
2HDM~\cite{Gunion:2002zf,Ginzburg:2004vp,Dev:2014yca,Pilaftsis:2016erj,Grzadkowski:2016szj},
which is favoured in the light of global analyses of experimental
constraints~\cite{Gorbahn:2015gxa,Belusca-Maito:2016dqe}.  In the
Higgs basis, the dimension-5 effective Lagrangian reads
\begin{equation}
-\mathcal{L}_{\rm eff}^{d=5}\ =\ \frac{y^2 }{M_{D}}
\frac{t_{\beta}}{1+t_{\beta}^2} \left( H_1^{\dagger}H_1-
  H_2^{\dagger}H_2 - t_{\beta} \, H_1^{\dagger}H_2 
+ t_{\beta}^{-1} \, H_2^{\dagger}H_1 \right) SS \  +\ \mathrm{H.c.}\;,
\label{HHSS}
\end{equation} 
where $H_1$ is the SM Higgs doublet and $H_2$ is the heavy scalar
doublet with $\langle H_2 \rangle=0$.

\subsection{Boltzmann Equation for {\boldmath $Y_S$}}
\label{eq:Boltzmann}

In order to determine the relic abundance of $S$ particles, we need to
solve the Boltzmann equation for their yield~$Y_S$. Since we assume
that the singlets $S$ remained out of equilibrium throughout the history
of the Universe (at least up to the phase of reheating), our only concern
will then be their production.  The main production channels, depending on
the plasma temperature $T$, are the following:
\begin{align}
    \label{channels}
 H_1^{\dagger}H_1,\ H_1^{\dagger}H_2,\ H_2^{\dagger}H_1\ \to\ & SS
  \qquad\qquad  \mbox{for }\  T_C \leq T < T_{\rm RH} \;, \nonumber \\
H_2^{\dagger}H_2\ \to\ & SS  \qquad\qquad\mbox{for }\  T < T_{\rm
  RH} \;, \nonumber\\
h\ \to\ & SS \qquad\qquad\mbox{for }\  T < T_C\;,
\end{align} 
where $h$ is the Higgs field with mass $m_h \approx 125~\mathrm{GeV}$
and $T_C \approx 130$~GeV is the critical temperature of the SM
electroweak phase transition. 
For $T<T_C$, one has to add new channels,  
for instance $W^{+}W^{-} \to SS$, but their contribution to
the production of the DM particles is negligible compared to
$h\to SS$.

Following~\cite{Hall:2009bx}, the Boltzmann equation for
the yield~$Y_{S}$ becomes
\begin{align}
sH\,\frac{dY_{S}}{dT} \ =\ & -\, \frac{1}{512
  \pi^5}\displaystyle\sum_{i,j=H_{1},H_{2}}\Bigg[\int_{(m_{i}+m_{j})^2}^{\infty}d\hat{s}\: 
  P_{ij} \: |M_{ij}|^2 K_{1}\bigg( \dfrac{\sqrt{\hat{s}}}{T}\bigg)\Bigg] \nonumber\\ 
 & +\, \bigg(\frac{t_{\beta}}{1+t_{\beta}^2 }\bigg)^2 \frac{y^4}{2\pi^3}
  \frac{ m_{h}^3 v^2}{M_{D}^2} K_{1}\bigg( \frac{m_h}{T} \bigg)\;.
\label{boltz}
\end{align}
where $T$ is the temperature of the plasma, $H$ is the Hubble
parameter, $K_{1}$ is the first modified Bessel function of the second
kind,
$P_{ij}\equiv
\sqrt{\hat{s}-(m_{i}+m_{j})^2}\sqrt{\hat{s}-(m_{i}-m_{j})^2}/\sqrt{\hat{s}}$
is a kinematic factor, and $|M_{ij}|^2$ is the squared matrix element,
summed over internal degrees of freedom, for the $2\to 2$ annihilation
processes: $H_i^\dagger H_j \to SS$.  The last term on the RHS
of~\eqref{boltz} arises from the decay $h \to SS$, upon ignoring the
mass of the $S$ particles. Also, upon ignoring $M_S$, the squared
matrix elements $|M_{ij}|^2$ for the various $2 \to 2$ processes are
\begin{align}
|M_{H_1^{\dagger}H_1 \to SS}|^2\ =\ & |M_{H_2^{\dagger}H_2 \to SS}|^2\
                                      =\  16
                                      \bigg(\frac{t_{\beta}}{1+t_{\beta}^2
                                      }\bigg)^2 y^4
                                      \frac{\hat{s}}{M_{D}^2}\; ,
                                      \nonumber \\ 
|M_{H_1^{\dagger}H_2 \to SS}|^2\ =\ & |M_{H_2^{\dagger}H_1 \to SS}|^2\
                                      =\ 8 (t_{\beta}^2 +
                                      t_{\beta}^{-2})
                                      \bigg(\frac{t_{\beta}}{1+t_{\beta}^2
                                      }\bigg)^2 y^4
                                      \frac{\hat{s}}{M_{D}^2}\;. 
\label{M^2}
\end{align}
The solution to the Boltzmann equation is obtained by integrating
\eqref{boltz} over the temperature~$T$. The limits of integration for the
various channels are the ones shown in \eqref{channels}.  However, before
doing that, we have to make an assumption for the critical temperature
and the thermal corrections to the masses of the scalar fields. In what
follows, we assume that the critical temperature~$T_C$ and the thermal
effects on the masses (for $T>T_C$) are similar to the pure SM
Higgs sector and they are given by~\cite{Fodor:1994bs}
\begin{align}
T_C \, \sim\,  m_h \;, \qquad
m_{H_1}^2\,  \approx\, m_{h}^2+ \dfrac{1}{2}T^2\;, \qquad  
m_{H_2}^2 \, \approx\,  \frac{1+t_{\beta}^2}{t_{\beta} } m_{12}^2  + \dfrac{1}{2}T^2\;.
\label{Thermal_masses}
\end{align}
Under these assumptions and restricting $T_{\rm RH}$ to be above $T_C$, we
can compute the yield~$Y_S$ at $T\approx 0$,  which in turn implies the relic
abundance~\cite{Edsjo:1997bg}
\begin{equation}
\Omega_S h^2\ \approx\ 0.12\times \bigg(\frac{M_S}{1~\mathrm{GeV}}\bigg) \, 
\bigg(\frac{Y_{S}(T=0)}{4.3\times 10^{-10}}\bigg)\;.
\label{Omega}
\end{equation}

\subsection{Approximate Results for {\boldmath  $\Omega_S h^2$}}  

In general, the yield~$Y_S$ cannot be calculated analytically, but
depending on the reheating temperature $T_{\rm RH}$, we are able to
present approximate analytic results.  We find that for
decoupled $D_{1,2}$, i.e. $T_{\rm RH} > M_D$, the relic abundance
$\Omega_S h^2$ derived from~$Y_S$ in \eqref{Omega} takes on the
form
\begin{equation}
    \label{Omega_approx_DD} 
\Omega_S h^2 \approx\ 0.12\times \bigg(\dfrac{M_S}{10^{-5}\, \mathrm{GeV}}\bigg)
\bigg(\dfrac{2 \times 10^{8}\, \mathrm{GeV}}{M_D}\bigg)^2 \bigg(\dfrac{y}{4.7
\times 10^{-2}}\bigg)^4 \bigg[ \bigg(\dfrac{t_{\beta}}{1+t_{\beta}^2}\bigg)^2
+ \bigg(\dfrac{T_{\rm RH}}{10^4 \, \mathrm{GeV}}\bigg)\bigg]\,,
\end{equation}
for
$T_{\rm RH}\gg m_{12}$, and
\begin{equation}
    \label{Omega_approx_AD}
\Omega_S h^2 \approx\ 0.12\times \bigg(\dfrac{M_S}{10^{-3}\,\mathrm{GeV}}\bigg)
\bigg(\dfrac{2 \times 10^{5}\, \mathrm{GeV}}{M_D}\bigg)^2 \bigg(\dfrac{y}{4.7
\times 10^{-4}}\bigg)^4 \bigg(\dfrac{t_{\beta}}{1+t_{\beta}^2}\bigg)^2 \bigg[
1 + \bigg(\dfrac{T_{\rm RH}}{10^4\, \mathrm{GeV}}\bigg) \bigg] \;,
\end{equation}
for $T_{\rm RH}\ll m_{12}$. Equations~(\ref{Omega_approx_DD} and~\ref{Omega_approx_AD}) 
are accurate up to $1\%$, except for $T_{\rm RH} \sim m_{H_2}$, where the deviation from the exact 
result is about $20\%$.
Note that in both the regimes of
$T_{\rm RH}$, there are two contributions to $\Omega_S h^2$, given by
the two terms contained in the last factors of~\eqref{Omega_approx_DD}
and~\eqref{Omega_approx_AD}.  The first contribution does not depend on the
reheating temperature~$T_{\rm RH}$ and arises from the decay
$h \to SS$, while the second one is proportional to~$T_{\rm RH}$.
This second contribution is a result of the decoupling of the heavy fermionic
doublets $D_{1,2}$ and indicates that for
$T_{\rm RH} \stackrel{>}{{}_\sim} 10^4~\mathrm{GeV}$, the production
of~$S$ particles is dominated by $2\to 2$ annihilation processes given
in~\eqref{channels}.  As discussed
in~\cite{Hall:2009bx,Elahi:2014fsa}, the latter is a general result
for the freeze-in production mechanism via non-renormalizable
operators.  Finally, it is worth pointing out that $\Omega_S h^2$ is
symmetric under $t_{\beta} \to t_{\beta}^{-1}$, as is the expression
for $M_S$ in \eqref{MS-1loop}.

\section{Results}
\label{sec:results}

In Section~\ref{sec:radmass}, we have shown that the mass of the singlet
$S$ can be generated at the one-loop level, if the PQ symmetry is softly
broken, and in Section~\ref{sec:DMabundance} we have calculated the relic
abundance of the $S$ particles.  In this section, we will be exploring
the validity of the parameter space of our minimal model.  To this end,
one may consider the parameters,
\[T_{\rm RH}\;,\, M_D\;, \,y^2 \;, \, t_{\beta}\; \text{ and }\;
m_{12}\;, \]
as being independent.  However, we prefer to solve the mass formula
$M_S^{\rm rad}$ in~\eqref{MS-1loop} for $y^2$ and replace it with a
physical observable, the $S$-particle mass $M_S$ which is taken in our
numerical estimates to be in the region:
$10~\mathrm{keV} \leq M_S \leq 1~\mathrm{GeV}$.  Consequently, the
parameters that we allow to vary independently are 
\begin{equation}
   \label{eq:inputs}
T_{\rm RH}\;, \, M_D\;, \,M_S\;, \, t_{\beta}\; \text{ and }\;
m_{12}\;. 
\end{equation}
We perform a scan over this parameter space, while imposing the
perturbativity constraint on the Yukawa couplings:~$Y_{1,2} <
\sqrt{4\pi}$. In this way, we find the values of these parameters
that satisfy the observed DM relic abundance~\cite{Ade:2013zuv}:
\begin{equation}
\Omega_S h^2\ =\ \Omega_{\rm DM} h^2\ =\ 0.1198 \pm 0.0026 \;.
\label{eq:relab}
\end{equation}

\begin{figure}[H]
\centering
        \includegraphics[width=0.72\textwidth]{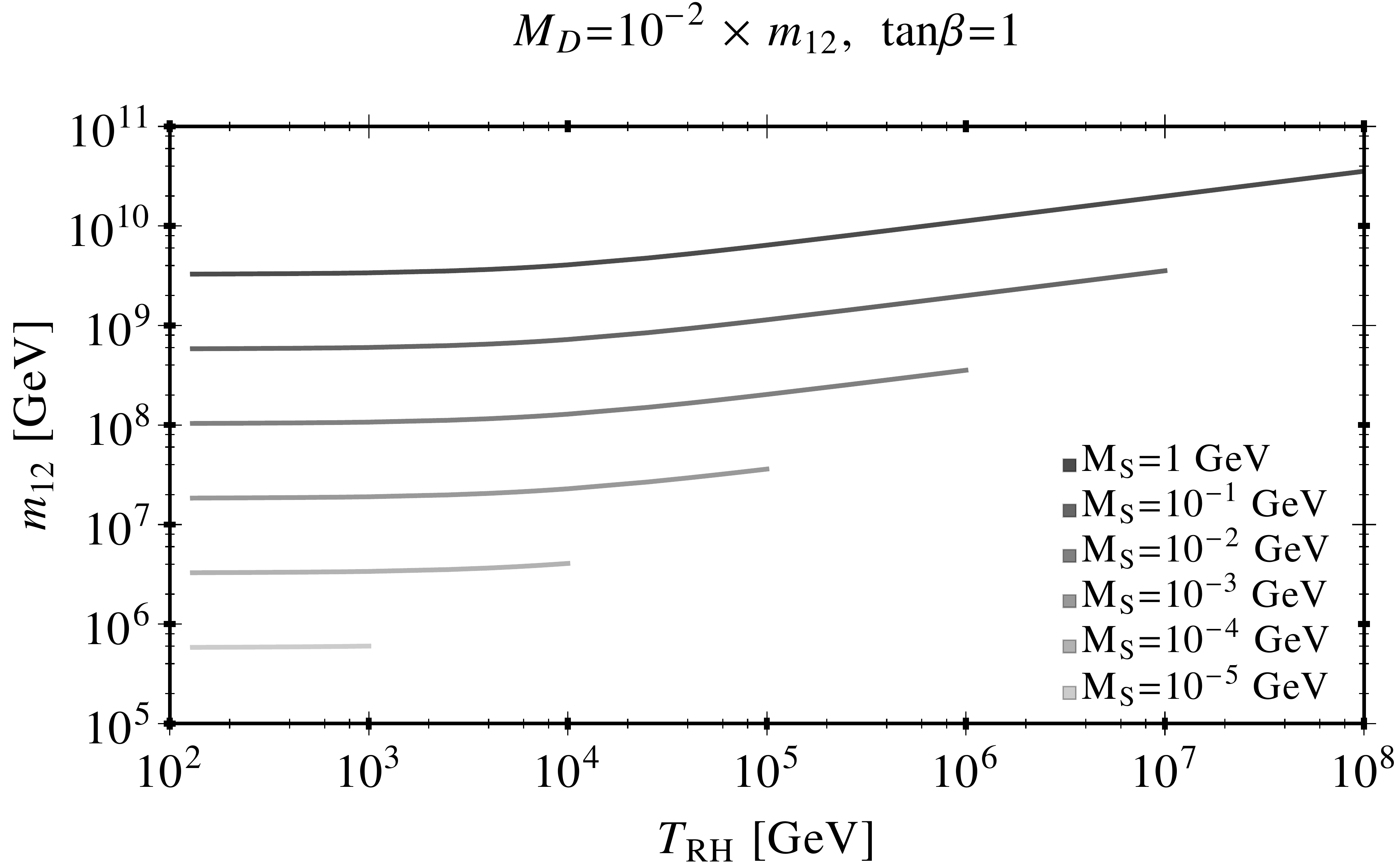}
        \caption{\sf $T_{\rm RH}$ versus $m_{12}$ for $r= M_D/m_{12} =10^{-2}$,
          $t_\beta=1$ and several RLDM masses~$M_S$.}
\label{TRH_vs_m12_gt_MD}
\end{figure}

\begin{figure}[H]
\centering
        \includegraphics[width=0.72\textwidth]{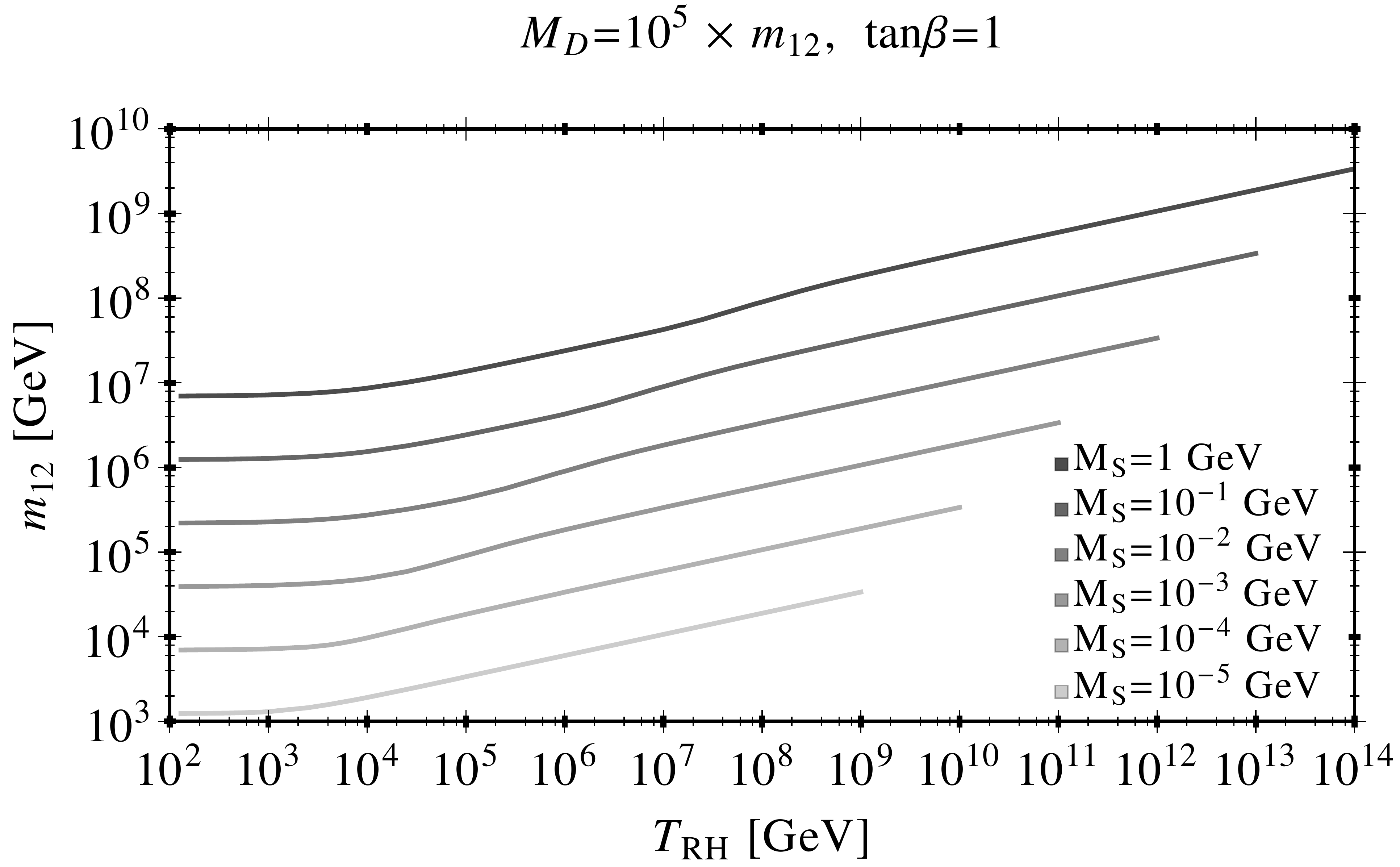}
\caption{\sf The same as in Figure~\ref{TRH_vs_m12_gt_MD}, but for $r=
  M_D/m_{12} = 10^5$.}
\label{TRH_vs_m12_lt_MD}
\end{figure}

In Fig.~\ref{TRH_vs_m12_gt_MD} we present contour lines on the
$T_{\rm RH}$--$m_{12}$ plane for discrete values of the $S$-particle
mass~$M_S$ in the region: $1~\mathrm{keV} \le M_S \le 1\GeV$, for
$t_{\beta}=1$ and $r=10^{-2}$, which give the DM relic
abundance~\eqref{eq:relab}.
For $m_{12} \simeq 10^{10}\GeV$, the reheating temperature
$T_{\rm RH}$ can vary between the critical temperature
$T_C\simeq 130$~GeV and $10^{8}$~GeV.
This upper bound on $T_{\rm RH}$ may be as high as
$10^{14}\GeV$, if the parameter $r=M_D/m_{12}$ is increased to the
value $r=10^5$, as depicted in Fig.~\ref{TRH_vs_m12_lt_MD}. Yet, at
the same time, $m_{12}$ increases by one order of magnitude or so.  On
the other hand, for $m_{12}\simeq 1-10\TeV$, an acceptable DM relic
abundance is reached only for large $r$ and for
$M_S \simeq 1~\mathrm{keV}$, as can be seen
from~Fig.~\ref{TRH_vs_m12_lt_MD}.
Most remarkably, we notice that the predicted values for
$\Omega_S h^2$ are compatible with the observed DM relic
abundance~$\Omega_{\rm DM} h^2$, for a wide range of values for the
parameters $m_{12}$,
$M_D$
and $T_{\rm
  RH}$.  Interestingly enough, the required Yukawa couplings~$Y_{1,2}$ for a
viable RLDM are sizeable, and always larger than the electron Yukawa
coupling.

We recall here that we explore  only regions where the fermion doublets $D_{1,2}$
are decoupled after the reheating of the Universe, i.e. we assume  $M_{D} \gg T_{\rm RH}$.
As a working hypothesis, we assume the decoupling condition:
$M_{D}> 3\, T_{\rm RH}$.  This condition is motivated by the fact for
$T \approx M_{D}/3$, the $D_{1,2}$ particles become non-relativistic and, as a
consequence, its number density is exponentially suppressed by a
Boltzmann factor.
Correspondingly, for the scenario considered in Fig.~\ref{TRH_vs_m12_gt_MD}, the heavy
scalar $H_{2}$ will be also decoupled, because $m_{H_2}\gg M_D$. 

Furthermore, we observe that for $T_{\rm RH}\stackrel{>}{{}_\sim} 10^4 \GeV$,
$m_{12}$ becomes linearly dependent on the reheating temperature, 
as expected from the approximate analytic expression in~\eqref{Omega_approx_AD}.
We also obtain a similar behaviour in Fig.~\ref{TRH_vs_m12_lt_MD}.  In
this case, however, the heavy scalar doublet~$H_2$ is no longer
constrained to be decoupled.  As a result, there is an interface
region at $T_{\rm RH}\sim m_{12}$ that lies between the two linear
regimes, $T_{\rm RH} \ll m_{12} $ and $T_{\rm RH}\gg m_{12}$.  At the
interface region, there is a transition caused by the contribution of
the heavy scalar doublet~$H_2$ to the production of singlet
fermions~$S$ [cf.~\eqref{channels}], which can reach equilibrium with
the plasma when~$T_{\rm RH}\gg m_{12}$.

\subsection{Solving the Strong CP Problem}

It is known that in the SM there is an explicit breaking of CP (and P) discrete
symmetry due to the instanton-induced term
\begin{equation}
{\cal L}_\theta\ =\ \frac{\theta}{32 \pi^2}\, \mathrm{Tr}(G_{\mu\nu}
\widetilde{G}^{\mu\nu})\;.
\end{equation}
In the above, $\theta$ is a CP-odd parameter which can be absorbed
into the quark masses. However, this $\theta$-parameter cannot be
fully eliminated, since  the combination:
$\bar{\theta} = \theta - \mathrm{Arg} \mathrm{Det} M_q$, where $M_q$
is the quark mass matrix, becomes a physical observable. It
contributes to the neutron dipole moment and experimentally, it is
severely bounded to
be:~$|\bar{\theta}| \stackrel{<}{{}_\sim}
10^{-11}$~\cite{Kim:2008hd}.
The problem of why $\bar{\theta}$ is much smaller than all other
CP-violating parameters, such as the well-known parameter
$\varepsilon_K \sim 10^{-3}$ from the $K^0\overline{K}^0$ system,
introduces another hierarchy problem in the SM known as the {\em
  strong CP problem}.  A possible solution, suggested by Peccei and
Quinn~\cite{Peccei:1977hh,Peccei:1977ur}, is to promote the
$\theta$-parameter into a dynamical field which naturally minimizes
the energy.  This dynamical field, called {\em the
  axion}~\cite{Weinberg:1977ma,Wilczek:1977pj}, is a pseudo-Goldstone
boson of the global anomalous PQ symmetry.

The SM has no global anomalous $U(1)_{\rm PQ}$-symmetry. One
possible way to realise such a symmetry is to non-trivially extend its
Higgs sector by adding a second Higgs doublet, resulting in the PQ-symmetric
2HDM. However, charging simply the field doublets $\Phi_1$ and
$\Phi_2$ under the PQ symmetry as done in Table~\ref{charges} does not
lead to a healthy model. Such a model predicts a visible keV-axion
with PQ-breaking scale $f_{\rm PQ} \sim 100\GeV$, which is
already excluded by the experiment. A minimal extension suggested by
Dine--Fischler--Sredniki\cite{Dine:1981rt}--Zhitnitsky~\cite{Zhitnitsky:1980tq}
(DFSZ) is to add a SM singlet $\Sigma$ with charge +1 under
$U(1)_{\rm PQ}$-symmetry such that the scalar potential term,

\begin{equation}
\lambda_\Sigma\,\Sigma^{2}\: \Phi_1^\dagger\:  \Phi_2  \ + \ \mathrm{H.c.}\
\subset\ V(\Phi_1,\Phi_2,\Sigma )\;,
\end{equation}
  is invariant.
Then, such a $\Sigma$-dependent term that occurs in the DFSZ potential
$V(\Phi_1,\Phi_2,\Sigma )$  breaks the PQ
symmetry spontaneously, when the electroweak singlet field $\Sigma$
receives a large VEV~$\langle \Sigma \rangle$ which is not necessarily tied in
with that of the electroweak scale~$v$.  For this reason, in this paper we have made the
identification 
\begin{equation}
\langle \Sigma \rangle\, \equiv\, f_{\rm PQ}\, \approx\, m_{12}\;,
\end{equation}
with $\lambda_\Sigma  \approx  1$. 
From experimental constraints and astrophysical considerations, the
PQ-breaking scale $f_{\rm PQ}$ must be typically larger than
$10^9 \GeV$~\cite{Raffelt:2006cw}.  Interestingly, within the RLDM
scenario, there are values for $m_{12}$ satisfying this constraint and
at the same time are compatible with the observed
$\Omega_{\rm DM} h^2$, as discussed in the previous section.  An
example is shown in Fig.~\ref{TRH_vs_m12_gt_MD} for $M_S = 1\GeV$ and
$T_{\rm RH}\ll m_{12}$.  In addition, values where
$m_{12} \stackrel{>}{{}_\sim} 10^9 \GeV$ can be also obtained for
other hierarchies e.g.~$r \sim 1$ and $r\gg 1$, as shown in
Fig~\ref{TRH_vs_m12_lt_MD}.  This seems to be a rather generic feature
of the RLDM realization.

Although the above is a strong indication that the DFSZ solution to
the strong CP problem is consistent with the RLDM scenario, a detailed
analysis of the UV-complete DSFZ-extended model lies beyond the scope of this
article. In particular, for
$f_{\rm PQ}\sim 10^{11} \GeV$~\cite{Turner:1985si}, the axion becomes
a sizeable DM component resulting in a two-component DM, consisting of
the axion and the $S$ particle, and so a more careful treatment will be required.

\subsection{Detection of RLDM}

We observe that for small enough reheating temperatures,
$T_{\rm RH}\sim 1~\mathrm{TeV}$, the fermion doublets $D_{1,2}$, as
well as the heavy scalar doublet $H_2$, can lie at the TeV scale,
provided that $M_S$ is of order $\mathcal{O}$(10~keV).  This is shown
in Figs.~\ref{TRH_vs_m12_gt_MD} and \ref{TRH_vs_m12_lt_MD} for light
$M_S$, where $M_D$ and $m_{12}$ lie in the vicinity of the TeV scale.
As a result, the DM particle $S$ can be probed indirectly by looking
for its associated ``partners'' of the heavy Higgs doublet~$H_2$. In general, we expect that
at the LHC, the heavy sector of the 2HDM will be efficiently explored
up to the TeV scale~\cite{Dev:2014yca,Pedersen:2016kyw}. For the RLDM
scenario at hand, however, such exploration may be somehow challenging, when
looking for charged Higgs bosons with masses larger than $\sim 1\TeV$
for a wide range of $t_{\beta}$
values~\cite{Dev:2014yca,Pedersen:2016kyw}.
 
On the other hand, direct detection experiments for sub-GeV DM
particles focus on their interactions with atomic electrons,
e.g.~see~\cite{Dedes:2009bk,Essig:2017kqs}.  However, in the RLDM
scenario, such a detection of $S$ particles is practically unattainable,
because $S$ interacts feebly with the SM Higgs boson with a coupling
proportional to $v/M_{D}\ll 1$ yielding a cross section for
$S \, e \to S \, e$, which is highly suppressed by fourth powers of
the electron-to-Higgs-mass ratio, i.e.
\begin{equation}
\bar{\sigma}_{Se}\ \approx\ 
\frac{y^4}{\pi} \frac{t_{\beta}^2}{(1+t_{\beta}^2)^2} \left( \frac{m_e}{m_h} \right)^4 \frac{1}{M_{D}^2}\ \approx \  
10^{-50} \times \frac{y^4 \, t_{\beta}^2}{(1+t_{\beta}^2)^2}\:
\bigg(\frac{1~\GeV}{M_{D}}\bigg)^2 \: \mathrm{cm}^2\;.
\end{equation}
Hence, a simple estimate shows that $\bar{\sigma}_{Se}$ is much smaller than its current experimental
reach:~$\bar{\sigma}^{\rm exp}_{Se} \simeq 10^{-38} \, \mathrm{cm}^2$.

Another potentially observable effect could originate from the invisible Higgs boson
decay, $h \to SS$.  Current LHC analyses report the upper
bound~\cite{Aad:2015txa}: $\mathrm{Br}(h \to \text{ inv.}) < 0.28$,
which for the RLDM scenario translates into
$$ M_D \ \stackrel{>}{{}_\sim}\  10^4 \times y^2 \frac{t_\beta}{1+t_\beta^2} \GeV\;.$$
Note that this constraint is comfortably satisfied for the entire range of our parameter space.

In summary, at least for the foreseeable future, the RLDM particle $S$
proposed here will remain elusive. This leaves only a window for the
LHC to find indirectly a second heavy Higgs doublet $H_2$ and/or a
pair of heavy fermion doublets~$D_{1,2}$.

\section{Conclusions}
\label{sec:Concl}

One central problem of most electroweak scenarios that require the
existence of very light DM particles in the keV-to-GeV mass range is the
actual origin of this sub-GeV scale. To~address the origin of such a
small scale, we have presented a novel radiative mechanism that can
naturally generate a sub-GeV mass for a light singlet fermion $S$,
which is stable and can successfully play the role of the DM.

In order to minimally realize such a Radiative Light Dark Matter, we
have considered a Peccei--Quinn symmetric two-Higgs doublet model,
which was extended with the addition of a singlet fermion $S$ and a
pair of massive vector-like SU(2) isodoublets $D_{1,2}$ that are not
charged under the PQ symmetry. Instead, the singlet fermion $S$ is
charged under the PQ symmetry and so it has no bare mass at the tree
level.  However, upon soft breaking of the PQ symmetry, we have shown
how the singlet fermion~$S$ receives a non-zero mass at the one-loop
level.
The so-generated
radiative mass for the singlet fermion~$S$ lies naturally in the
cosmologically allowed region of $\sim 10$~keV--1~GeV.

We have computed the relic abundance of the RLDM $S$, for different
plausible heavy mass scenarios. Specifically, for all scenarios we
have been studying, we
have assumed that the $S$ particles were absent in the early Universe,
whilst the fermion isodoublets~$D_{1,2}$ stay out of equilibrium
through the entire thermal history of the Universe, because their
gauge-invariant mass $M_D$ is taken to be well above the reheating
temperature~$T_{\rm RH}$.  Then, we have found that the
observationally required relic abundance for the RLDM $S$ can be
produced via decays and annihilations of Higgs-sector particles.

We have analyzed a heavy mass scenario where the PQ-breaking
scale~$f_{\rm PQ}$ can
reach values $\sim 10^9$~GeV as required by the
Dine--Fischler--Sredniki--Zhitnitsky axion model to explain the strong
CP problem. We have found that for appropriate isodoublet masses
(e.g. in Fig.~\ref{TRH_vs_m12_gt_MD} $M_D \sim 10^{-2}\,f_{\rm PQ}$), the RLDM particle $S$ in such a scenario can
successfully account for the missing matter component of the
Universe. In addition, we have investigated whether a lower mass limit
exists for the heavy Higgs scalars, within the context of a viable
RLDM scenario. We have found that the masses of the heavy scalars can
be as low as TeV, which allows for their possible detection at the
LHC in the near future.

The PQ-symmetric scenario we have studied here generates a viable RLDM
at the one-loop level. However, one may envisage other extensions of
the SM, in which the required small mass for the light DM could be
produced at two or higher loops. For instance, if the SM is
extended by two scalar triplets, a small DM mass can be
generated through their mixing at the two-loop level, in a fashion similar to the Zee
model. In this context, it would be interesting to explore possible
models where both the tiny mass of the SM neutrinos and the small mass
of the light DM have a common radiative origin and study their
phenomenological implications.

\subsection*{Acknowledgements}

The work of AP is supported in part by the
Lancaster--Manchester--Sheffield Consortium for Fundamental Physics,
under STFC research grant ST/L000520/1.

\newpage

\newpage
\renewcommand{\thesection}{Appendix~\Alph{section}}
\renewcommand{\theequation}{\Alph{section}.\arabic{equation}}

\setcounter{equation}{0}  
\setcounter{section}{0}
\bigskip


\bibliography{RDM.bib}{}
\bibliographystyle{JHEP}

\end{document}